\documentclass[aps,twocolumn,showpacs]{revtex4}
\usepackage{graphics}
\usepackage{graphicx}

\begin{document}
\title{Depletion of carriers and negative differential conductivity \\
in an intrinsic graphene under a dc electric field}
\author{P.N. Romanets}
\author{F.T. Vasko}
\email{ftvasko@yahoo.com}
\affiliation{Institute of Semiconductor Physics, NAS of Ukraine,
Pr. Nauky 41, Kiev, 03028, Ukraine}
\date{\today}

\begin{abstract}
The heating of carriers in an intrinsic graphene under an abrupt switching
off a dc electric field is examined taking into account both the energy
relaxation via acoustic and optic phonons and the interband generation-recombination processes. The later are caused by the interband transitions
due to optical phonon modes and thermal radiation. Description of the temporal
and steady-state responses, including the nonequilibrium concentration and
energy as well as the current-voltage characteristics, is performed. At
room temperature, a nearly-linear current-voltage characteristic and a
slowly-varied concentration take place for fields up to $\sim$20 kV/cm.
Since a predominant recombination of high-energy carriers due to optical
phonon emission at low temperatures, a depletion of concentration takes
place below $\sim$250 K. For lower temperatures the current tends to be
saturated and a negative differential conductivity appears below $\sim$170 K
in the region of fields $\sim$10 V/cm.
\end{abstract}


\maketitle

\section{Introduction}
In addition to detailed investigation of the linear transport phenomena
in graphene, see reviews \cite{1} and references therein, a high-field
transport regime is coming under examination during recent years. Starting
from the first experiments on current cleaning of graphene, \cite{2} a set
of measurements on transistor structures, where lateral inhomogeneities
of concentration and contact phenomena are essential, were performed, see
last results and reviews. \cite{3,4} Hot electron transport in short-channel suspended graphene devices was studied in Ref. 5. Recently, a few
measurements \cite{6,7} were performed on homogeneous structures without
contacts contribution. As for theoretical description, the energy
relaxation processes in graphene were studied in several papers. \cite{8}
Heating of carriers by a strong dc electric field was considered both
analytically \cite{9,10} and numerically, with the use of the Monte
Carlo method. \cite{11} Most of these results \cite{2,3,4,5,6}
and \cite{8,10,11} were obtained for the monopolar graphene, with a
fixed concentration of electrons (or holes). For the case of intrinsic
graphene, not only the energy (temperature) of carriers increases due
to the Joule heating but also {\it a carrier concentration changes}. \cite{9}
Thus, the nonequilibrium distribution is determined both the energy
relaxation and the interband generation-recombination processes. This
regime of heating was analyzed in \cite{9} for the low-energy carriers,
at electric fields $E\leq$10 V/cm and at low temperatures. The
modifications of the current-voltage characteristics, $I(V)$ (here $V$
is a drop of voltage along structure), with variation of the gate
voltage, $V_g$, from the heavily-doped regime of transport to the
intrinsic case (at $V_g=$0), were reported in \cite{7} for the room
temperature. While the mechanism for formation of the second Ohmic
law in $I(V)$ under the monopolar regime of transport was considered
in \cite{7}, an appearance of the quasilinear characteristic at $V_g=$0,
when the generation-recombination processes are essential, remains unclear.

In this paper, we consider a heating of carriers after an abrupt switching
off a dc electric field. In addition to the quasielastic relaxation via
acoustic phonons and the interband transitions due to thermal radiation
considered in Ref. 9, the intra- and interband emission and absorption of
optical phonons is taken into account in the high-energy region. The Cauchy problem for the quasiclassical kinetic equation is solved below for the
case of a weakly anisotropic distribution under an effective momentum
relaxation. The temporal concentration and energy as well as the
current-voltage characteristics are analyzed and their dependencies on
temperature and field strength are described for the steady-state conditions.

The results obtained can be briefly summarized as follows. At room
temperature, {\it a nearly-linear} $I(V)$ characteristic and a
slowly-increased concentration dependency take place up to fields
$E\sim$20 kV/cm in agreement with the experimental data. \cite{7} This
is because an interplay between the emission and absorption of optical
phonons (the last process is proportional to the nonzero Planck number
of $\Gamma$- and $K$-modes). For lower temperatures, $I(V)$ tends
to be saturated (starting $E\sim$20 V/cm at $T<$150 K) and {\it a
negative differential conductivity} (NDC) appears in the transition
region ($E\sim 5$ - 15 V/cm at 77 K). Since a predominant recombination
of high-energy carriers due to optical phonon emission at low
temperatures, {\it a depletion of concentration} takes place below
250 K. The semi-insulating regime of conductivity with a residual
concentration about $10^8$ cm$^{-2}$, an energy per particle less
80 meV, and with a low saturated current is realized at 77 K and
$E\gg$20 V/cm. Beside of the peculiarities of the steady-state response,
{\it a two-scale temporal evolution} takes place due to a fast relaxation
via optical phonons and a slow relaxation via acoustic phononsand thermal
radiation. A scale of transient process is strongly dependent on temperature
and on fields applied. At low temperatures a steady-state distribution is established during $\sim$10 ns time scale and the transient response can
be measured directly.

The paper is organized as follows. The basic equations governing the
heating of carriers under an abrupt switching off a dc electric field
are considered in Sec. II. Temporal evolution of nonequilibrium distributions
for different temperatures and electric fields is described in Sec. III.
Description of the temporal and steady-state responses, including the nonequilibrium concentration and energy as well as the current-voltage characteristics, is presented in Sec. IV. The concluding remarks and
discussion of the assumptions used are given in the last section.

\section{Kinetic approach}
Nonequilibrium electrons and holes in an intrinsic graphene are described
by coinciding distributions $f_{{\bf p}t}$ because their energy spectra are symmetric and the scattering mechanisms are identical. \cite{9} An evolution
of these distributions under a homogeneous electric field ${\bf E}_t$ is
governed by the quasiclassical kinetic equation
\begin{equation}
\frac{\partial f_{{\bf p}t}}{\partial t}+e{\bf E}_t\cdot\frac{\partial
f_{{\bf p}t}}{\partial {\bf p}}=\sum\limits_j J_j (f_t |{\bf p}) .
\end{equation}
Here the collision integrals $J_j(f_t |{\bf p})$ describe the relaxation
of carriers caused by the elastic scattering on structure disorder ($j=d$),
the intraband quasielastic scattering on acoustic phonons ($j=qe$), and
the intra- and interband emission and absorption of optical phonons ($j=intra,~inter$). In addition, the contribution $J_r(f_t |{\bf p})$
describes the generation-recombination processes due to interband
transitions induced by thermal radiation. At $t<0$, we use the equilibrium
initial condition $f_{{\bf p}t<0}=f^{(eq)}_p$ at temperature $T$
where $f^{(eq)}_p=\{\exp (p/p_T)+1\}^{-1}$ and $p_T=T/\upsilon$ is
written through the characteristic velocity $\upsilon\simeq 10^8$
cm$^{-2}$.

For typical graphene structures, the momentum relaxation dominants over
the other processes listed, i.e. $\nu_d\gg\nu_j,~j\neq d$ where $\nu_j$
stands for the relaxation rate of the $j$-th scattering channel. Under the
condition $eE\tau_m\ll\overline{p}$, where $\tau_m\sim\nu_d^{-1} $ is the
momentum relaxation time and $\upsilon\overline{p}$ is the energy of hot
carriers, the anisotropy of distribution is weak, $f_{{\bf p}t}\simeq f_{pt}+\Delta f_{{\bf p}t}$. Here we separated the isotropic part of distribution,
$f_{pt}$, and a weak anisotropic contribution, $\Delta f_{{\bf p}t}$.
Within the local time approximation, when ${\bf E}_t$ increase slowly over
$\tau_m$-scale, we obtain the asymmetric part of distribution at $t>0$ as
\begin{equation}
\Delta f_{{\bf p}t}\simeq\frac{e{\bf E}\cdot {\bf p}}{p\nu_p}\left(
-\frac{\partial f_{pt}}{\partial p} \right) , ~~ \nu_p=\frac{v_d p}
{\hbar}\Psi\left(\frac{pl_c}{\hbar}\right) +\frac{v_0 p}{\hbar} .
\end{equation}
The phenomenological rate $\nu_p$ is written here through the characteristic velocities $v_d$ and $v_0$ for the case of the combined elastic scattering
by short- and long-range disorder, see Ref. 12 for details. For the model of
the Gaussian disorder with the correlation length $l_c$ we use the
dimensionless function $\Psi (z)=e^{-z^{2}}I_{1}(z^{2})/z^{2}$ with the
first-order Bessel function of an imaginary argument, $I_{1}(z)$.

Performing the averaging over in-plane angle (such averaging symbolized by overline) in kinetic equation (1) and neglecting the weak contribution of
$\Delta f_{{\bf p}t}$ in the right-hand side, one obtains the kinetic equation
for symmetric distribution $f_{pt}$ as follows
\begin{equation}
\frac{\partial f_{pt}}{\partial t}+ e\overline{{\bf E}\cdot\frac{\partial
\Delta f_{{\bf p}t}}{\partial{\bf p}}} =\sum_{j\neq d}J_j (f_t |p) .
\end{equation}
Because the elastic scattering does not affect the symmetric distribution due
to the energy conservation law, the only non-elastic mechanisms ($j\neq d$)
are responsible for relaxation of $f_{pt}$. The Joule heating contribution ($\propto E^2$ term in the left-hand side) is expressed through the asymmetric correction (2) and it can be transformed into
\begin{equation}
e\overline{{\bf E}\cdot\frac{\partial \Delta f_{{\bf p}t}}{\partial{\bf p}}}
=\frac{(eE)^2}{2p}\frac{\partial}{\partial p}\left[ \frac{p}{\nu_p}\left(
-\frac{\partial f_{pt}}{\partial p}\right)\right] .
\end{equation}
Eqs. (3, 4) should be solved at $t>0$ with the initial condition $f_{pt=0}
=f^{(eq)}_p$ and the use of the collision integrals described below.

Within the quasielastic approximation, \cite{9,13} the energy relaxation
via acoustics phonons is described by the Fokker-Planck collision integral
given by
\begin{eqnarray}
J_{qe}(f_t|p) = \frac{\nu _p^{(qe)}}{p^2}\frac{\partial}{\partial p}\left[
p^4\frac{\partial f_{pt}}{\partial p}+\frac{p^4}{p_T}f_{pt}(1-f_{pt}) \right] .
\end{eqnarray}
Here the rate $\nu _p^{(qe)}=(s/v)^2v_{ac}p/\hbar$ is written through the
sound velocity $s$ ($\upsilon /s\simeq$137) and the characteristic velocity
$v_{ac}\simeq$2.4$\times 10^5$ cm/s at the nitrogen temperature, moreover
$v_{ac}\propto T$. The interband transitions caused by the thermal radiation
are described by the collision integral
\begin{equation}
J_{R}(f_t|p) =\nu_p^{(R)}[N_{2p/p_T}(1 - 2f_{pt})-f_{pt}^2] ,
\end{equation}
where the rate of spontaneous radiative transitions, $\nu_p^{(R)}=v_rp/\hbar$,
is written through the characteristic velocity $v_r\simeq$41.6 cm/s for the
case of graphene on the SiO$_2$ substrate.

The scattering by optical phonons is described by the collision integral
written through the intra- and interband parts, $J_{intra}(f_t|p)+J_{inter}
(f_t|p)$, as follows (see evaluation in Refs. 14 and 15)
\begin{eqnarray}
J_{intra}(f_t|p)=\sum\limits_\eta\left[ (N_\eta +1)\nu_{p+p_\eta}^{(\eta )}
(1-f_{pt})f_{p+p_{\eta}t} \right. \nonumber \\
+N_{\eta}\nu_{p-p_\eta}^{(\eta )}(1-f_{pt})f_{p-p_{\eta}t}-(N_\eta +1)
\nu_{p-p_\eta}^{(\eta )} \\
\left.\times(1-f_{p-p_{\eta}t})f_{pt} -N_\eta\nu_{p+p_\eta}^{(\eta )}(1-f_{p+p_{\eta}t})f_{pt} \right]  \nonumber
\end{eqnarray}
and
\begin{eqnarray}
J_{inter}(f_t|p)=\sum\limits_\eta\left[ N_\eta\widetilde\nu_{p_\eta -p}^{(\eta )}
(1-f_{pt})(1-f_{p_\eta -pt}) \right. \nonumber \\
\left. -(N_\eta +1)\widetilde\nu_{p_\eta -p}^{(\eta )}
f_{p_\eta -pt} f_{pt} \right] .
\end{eqnarray}
Here $N_{\eta}=[\exp (\hbar\omega_{\eta}/T)-1]^{-1}$ is the Planck distribution
of $\eta$th phonon mode with the energy $\hbar\omega_{\eta}$ at temperature $T$ and $p_\eta =\hbar\omega_\eta /\upsilon$ is the characteristic momentum.
Summation over $\eta$ involves both the intra- and intervalley transitions
(marked by $\eta =\Gamma$ and $K$ respectively) taking into account the
zone-center and zone-boundary phonon modes. Since the averaging over in-plane angle in the general form of the collision integral, \cite{14,15} one obtains
the relaxation rates $\nu_p^{(\eta )}\simeq\theta (p)v_\eta p/\hbar$ and $\widetilde\nu_p^{(\eta )}\simeq\theta (p)\widetilde{v}_\eta p/\hbar$, which
are proportional to the density of states. The $\theta$-function here allows
the interband transitions, if $p<p_\eta$ (in the passive region). Below we
suppose the same relaxation rates for the intra- and intervalley transitions, $v_\eta\simeq\widetilde{v}_\eta$ and we use the characteristic velocities $v_\Gamma\simeq 10^6$ cm/s and $v_K\simeq 2\times 10^6$ cm/s which are in agreement of the previous results. \cite{14,15}

The concentration of electrons (holes), $n_t$, and the energy and current
densities, ${\cal E}_t$ and ${\bf I}_t$, are determined through $f_{pt}$ and
$\Delta f_{{\bf p}t}$ according to the standard relations:
\begin{equation}
\left|\begin{array}{*{20}c} n_t \\ {\cal E}_t \\ {\bf I}_t \end{array} \right|
=4\int\frac{d{\bf p}}{(2\pi\hbar )^2}\left| \begin{array}{*{20}c}
f_{pt} \\ \upsilon pf_{pt} \\ e{\bf v}_{\bf p}\Delta f_{{\bf p}t}\end{array}
\right| .
\end{equation}
The factor 4 here takes into account the spin and valley degeneracy and
${\bf v}_{\bf p}=\upsilon{\bf p}/p$ is the carrier velocity. Using the local
time approach, see Eq. (2), and introducing the nonlinear conductivity
$\sigma_t$ according to ${\bf I}_t=\sigma_t{\bf E}_t$, one obtains
\begin{equation}
\sigma_t =\frac{e^2\upsilon}{\pi\hbar^2}\int\limits_0^\infty\frac{dpp}{\nu _p}
\left( -\frac{\partial f_{pt}}{\partial p} \right) .
\end{equation}
For the case of the short-range scattering ($l_c\to$0 and $\overline{v}_d=
v_d+v_0$), when $\nu_p\simeq\overline{v}_dp/\hbar$, this integral is
transformed into $\sigma_t\simeq (e^2\upsilon /\pi\hbar^2\overline{v}_d)
f_{p=0t}$, i.e. $\sigma_t\propto f_t$ for the low-energy region [$\upsilon
p<$60 - 100 meV, see Fig. 1(a) below].
\begin{figure}[ht]
\begin{center}
\includegraphics[scale=1.2]{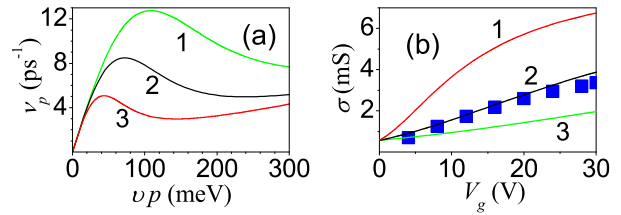}
\end{center}
\addvspace{-1 cm}
\caption{(Color online) (a) Momentum relaxation rate $\nu_p$ versus energy
$\upsilon p$ for $l_c=$5 nm (1) 7.5 nm (2), and 12.5 nm (3). (b) Conductivity
versus gate voltage for the same $l_c$ as in panel (a). Squares are experimental
points from Fig. 2a of Ref. 7. }
\end{figure}

\section{Transient evolution of distribution}
First, we consider numerical solution of the Cauchy problem for the nonlinear differential equation with the finite-difference terms given by Eqs. (4)-(8).
Using the iteration scheme \cite{16} at $t>0$ and the equilibrium initial condition one obtains the transient distribution $f_{pt}$ for different
electric fields and temperatures. Under these calculations we use the
momentum relaxation rate given by Eq. (2) at different correlation lengths,
$l_c=$5 - 12.5 nm, with the characteristic velocity $v_d\simeq 2.6\times
10^7$ cm/s, which is correspondent to the maximal sheet resistance
$\sim$3.3 k$\Omega$ and $v_0/v_d\simeq$0.035. In Figs. 1(a) and 1(b) we plot
the relaxation rate $\nu_p$ versus energy $\upsilon p$ and the linear
conductivity $\sigma$ versus gate voltage, $V_g$, for $l_c=$5, 7.5 and
12.5 nm. As it is shown in Fig. 1(b), the curve 2 is in agreement with the experimental data of Ref. 7 and below we use these parameters for description of the nonlinear response.
\begin{figure}[ht]
\begin{center}
\includegraphics[scale=1.2]{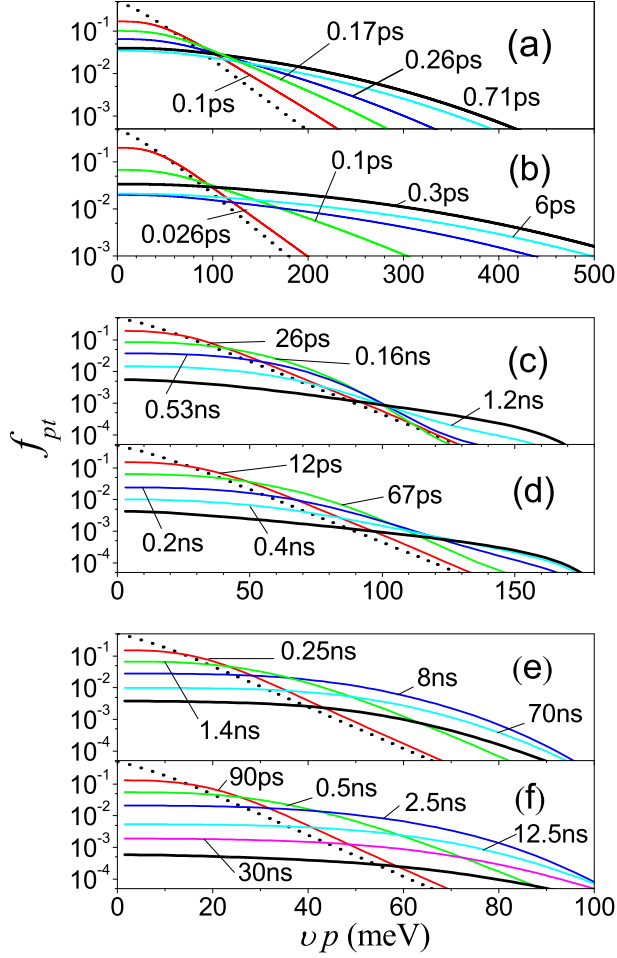}
\end{center}
\addvspace{-0.5 cm}
\caption{(Color online)  Distribution function $f_{pt}$ versus energy $vp$
for times marked at different temperatures and electric fields: (a) $T=$300 K
and $E=$6 kV/cm; (b) $T=$300 K and $E=$12 kV/cm; (c) $T=$150 K and  $E=$60 V/cm;
(d) $T=$150 K and  $E=$120 V/cm;  (e) $T=$77 K and  $E=$10 V/cm; (f) $T=$77 K
and and  $E=$20 V/cm. Dotted and solid black curves are correspondent to
initial and final distribution, $f^{(eq)}_p$ and $f_p$, respectively. }
\end{figure}

Transient evolution of distribution $f_{pt}$, from equilibrium form $f_p^{(eq)}$
to steady-state function $f_p\equiv f_{pt\to\infty}$, is shown in Figs.
2(a) - 2(f) for different temperatures and electric fields. At room temperature,
the transient process takes place during 1 - 10 ps time scale and this process becomes slower for higher fields because an interplay between emission and
absorption of optical phonons. At lower temperatures, the evolution times are
longer, about 1 ns at 150 K and 50 ns at 77 K, but these time scales become
shorter for higher fields in contrast to the room temperature case. It is because $N_\eta\leq 10^{-5}$ at $T<$200 K and the absorption of optical phonons is negligible. The maximal distribution $f_{p=0t}$ decreases during the transient process, moreover this effect is enhanced for higher fields and lower temperatures.
\begin{figure}[ht]
\begin{center}
\includegraphics[scale=1]{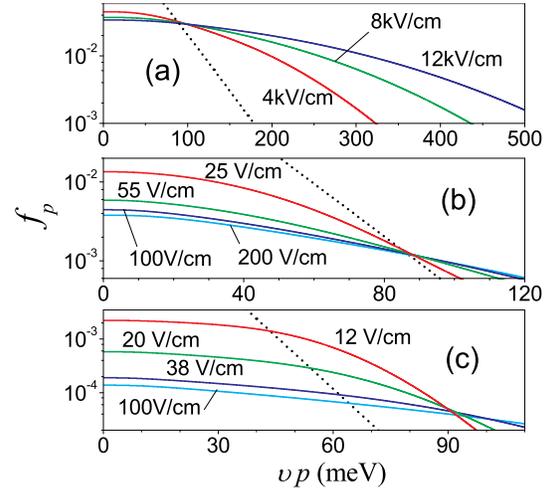}
\end{center}
\addvspace{-0.5 cm}
\caption{(Color online) (a)  Distribution function $f_p$ versus $vp$ for
$T=$300 K at different electric fields (marked). (b) The same for $T=$150 K.
(c) The same for $T=$77 K. Dotted lines are correspondent to the equilibrium
distribution. }
\end{figure}

The steady-state distribution functions, $f_p$, which are established after
the transient process, are shown in Fig. 3 for different field strengths and
$l_c=$7.5 nm at temperatures 300 K, 150 K, and 77 K. In contrast to the exponential decay of the equilibrium distribution at $E=0$ shown as the
dotted lines, at $E\neq 0$ one obtains a smeared distribution with a low
maximal value $f_{p=0}\leq 10^{-2}$. At room temperature, carriers are
distributed over energies $\gg\hbar\overline{\omega}_\eta$, if $E>$ 2 kV/cm
[see Fig. 3(a)] due to competition between absorption and emission of optical phonons. Here and below, $\hbar\overline{\omega}_\eta$ stands for the lowest
optical phonon energy. At low temperatures, the distributions are located
in the passive region of energies $\upsilon p<\hbar\overline{\omega}_\eta$
[see Figs. 3(b) and 3(c)] due to effective emission of optical phonons.
\begin{figure}[ht]
\begin{center}
\includegraphics[scale=1]{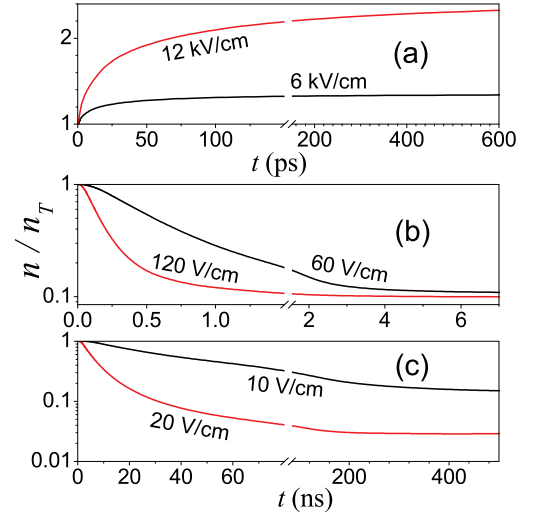}
\end{center}
\addvspace{-0.5 cm}
\caption{(Color online) Transient evolution of concentrations $n_t$
(normalized to their equilibrium values, $n_T$) for $l_c=$7.5 nm and
different electric fields (marked) at $T=$300 K (a), 150 K (b), and 77 K (c). }
\end{figure}

\section{Results}
The nonlinear responses introduced by Eqs. (9, 10) are analyzed below. Both
the transient characteristics and the steady-state dependencies of concentration,
energy and current densities on temperature and on dc field are described.

\subsection{Concentration and energy vs $t$ and $E$}
Transient evolution of concentrations $n_t$, normalized to the equilibrium
concentration in an intrinsic graphene [$n_T\simeq 0.52(T/\hbar\upsilon )^2$
varies between $8.1\times 10^{10}$ cm$^{-2}$ and $5.4\times 10^9$ cm$^{-2}$
for $T=$300 - 77 K] is shown in Figs. 4(a) - 4(c) for the same temperatures and electric fields as in Fig. 2. At room temperature, $n_t$ increases with
$t$ and the saturation regime is realized at longer times for higher fields
($t\geq$0.1 ns for $E=$12 kV/cm). It is  because the distribution is moved
away from the passive region [see Figs. (2a) and (2b)] where the generation-recombination processes take place. At lower temperatures, $n_t$ decreases
during temporal evolution and the saturation regime is realized for time
intervals $>$0.5 ns or $>$50 ns at $T=$150 K or 77 K, respectively. The
depletion of concentration is realized because an effective recombination
via the optical phonon emission appears at $\upsilon p>\hbar
\overline{\omega}_\eta /2$. All temporal dependencies appear to be faster
during initial stages of evolution at high fields, i. e. a two-stage
transient process takes place because an increasing of the relaxation rates
with energy.

\begin{figure}[ht]
\begin{center}
\includegraphics[scale=1]{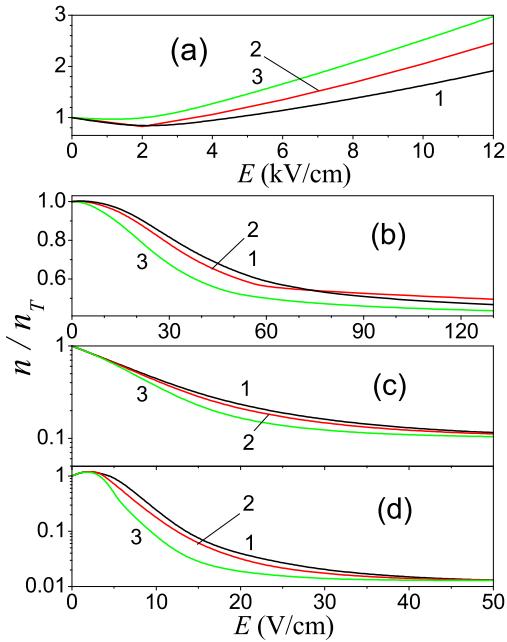}
\end{center}
\addvspace{-1 cm}
\caption{(Color online) (a) Concentrations of carriers $n$ (normalized to their
equilibrium values, $n_T$) versus electric field $E$ for $T=$300 K (a),
225 K (b), 150 K (c), and 77 K (d). Curves 1 - 3 are correspondent to $l_c=$5, 7.5, and 12.5 nm. }
\end{figure}
Variation of the normalized nonequilibrium concentration $n/n_T$ with field applied is shown in Figs. 5(a)-(d) for different temperatures and $l_c$. At
room temperature $n$ increases with $E$ due to the thermogeneration process
(first term of Eq. (8) with $N_K\sim 1.4\times 10^{-3}$). At lower temperatures, if $T<$250 K when $N_\eta <10^{-4}$, the thermogeneration is negligible and
$J_{inter}<0$. As a result, depletion of concentration takes place as $E$ increases. In a high-field region [at $E\geq$50 V/cm if $T=$77 K, see Fig.
5(d)], one arrives to the saturation regime with concentrations lower
$10^8$ cm$^{-2}$. The generation-recombination processes appears to be more effective if $l_c\geq$10 nm, so that $n$ decreases faster if $E$ increases.
\begin{figure}[ht]
\begin{center}
\includegraphics[scale=1]{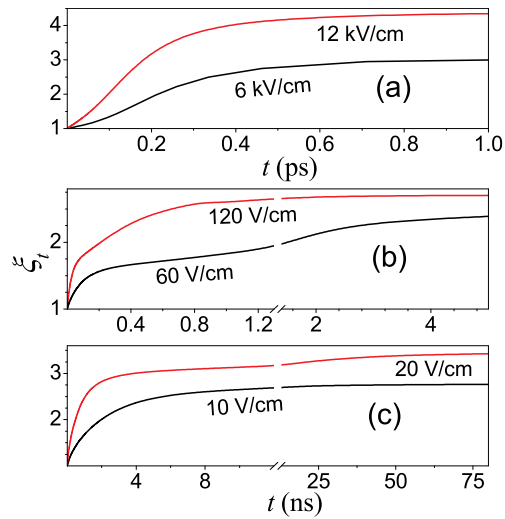}
\end{center}
\addvspace{-0.5 cm}
\caption{(Color online) Transient evolution of energy per carrier normalized
to their equilibrium values, $\xi_t$ given by Eq. (11) for $l_c=$7.5 nm and different electric fields (marked) at $T=$300 K (a), 150 K (b), and 77 K (c). }
\end{figure}

\begin{figure}[ht]
\begin{center}
\includegraphics[scale=1]{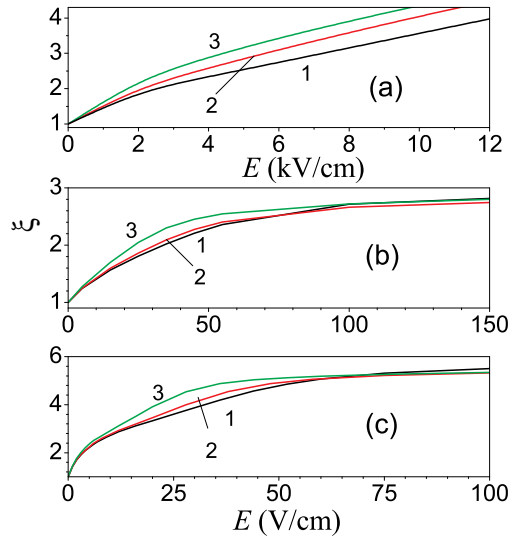}
\end{center}
\addvspace{-0.5 cm}
\caption{(Color online) The ratio $\xi =\xi_{t\to\infty}$ given by Eq. (11)
versus electric field $E$ for $T=$300 K (a), 150 K (b) and 77 K (c).
Curves 1 - 3 are correspondent to $l_c=$5, 7.5, and 12.5 nm. }
\end{figure}
It is convenient to characterize the energy distribution by the averaged
energy per carrier, ${\cal E}_t/n_t$, normalized to their equilibrium
value ${\cal E}_T/n_T$, i. e. we consider below the ratio
\begin{equation}
\xi_t =\frac{{\cal E}_t/n_t}{{\cal E}_T/n_T} ,
\end{equation}
where ${\cal E}_T/n_T\simeq 0.22T$. Transient evolution of $\xi_t$ is
plotted in Figs. 6(a) - 6(c) for the same $T$ and $E$ as in Figs. 2 and 4.
Notice, that the energy per carrier saturates faster in comparison to concentration, c. f. Figs. 6 and 4. The steady-state ratio $\xi =
\xi_{t\to\infty}$ is shown in Fig. 7 versus electric field for different
$l_c$ and temperatures. After a fast increasing of $\xi$ with $E$ at low
fields, one obtains a slow dependency of the energy per carrier versus $E$.
The dependencies on $l_c$ are similar to ones shown in Fig. 5.

\begin{figure}[ht]
\begin{center}
\includegraphics{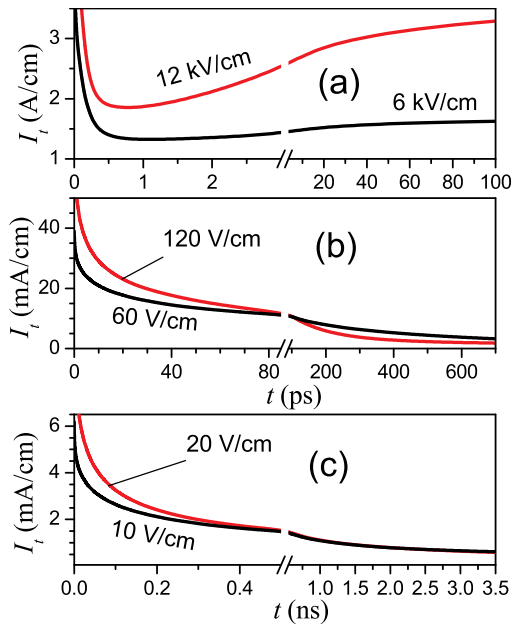}
\end{center}
\addvspace{-0.5 cm}
\caption{(Color online) Transient evolution of current-voltage characteristics
for the same conditions as in Fig. 4.  }
\end{figure}

\begin{figure}[ht]
\begin{center}
\includegraphics{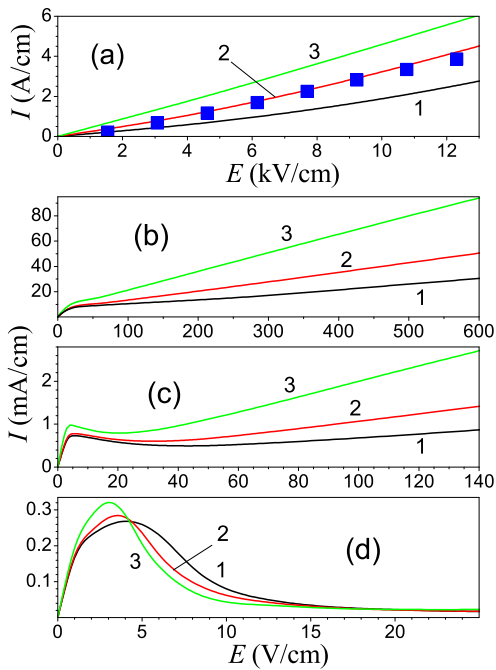}
\end{center}\addvspace{-1 cm}
\caption{(Color online) Current-voltage characteristics for the same
conditions as in Fig. 5. Experimental points from Fig. 2(b) of Ref. 7 are
shown in panel (a) as squares. }
\end{figure}
\subsection{Current-voltage characteristics}
We turn now to consideration of the current-voltage characteristics, $I_t$
versus $E_t$ normalized to length of sample. Transient evolution of current
$I_t$ is shown in Fig. 8 for the same $T$ and $E$ as in Figs. 2, 4, and 6.
Once again, one obtains the two-stage evolution over time scales similar to
those for the $n_t$ and $\xi_t$ dependencies. At low temperatures (150 K and
77 K) $I_t$ decreases monotonically, similarly to $n_t$. At room temperature,
$I_t$ decreases fast during an initial stage of the transient process
because energy and momentum relaxation rates increase. For longer times,
$I_t$ increases similarly to the $n_t$ dependency at $t\geq$5 ps, c. f.
Figs. 4(a) and 8(a).

The steady-state current-voltage characteristics, $I(E)$, are shown in
Fig. 9 under the same conditions as in Figs. 5 and 7. For the low-field
region (about tens V/cm at $T\geq$200 K and $\sim$V/cm at lower $T$), the
$I(E)$ dependencies are in agreement with the results reported in Ref. 9.
In the high-field region, the second Ohmic law with the resistance
$\sim$3 k$\Omega$ at $T=$300 K takes place, while the saturation of current
with $I\leq$10 $\mu$A/cm takes place at low temperatures (a semi-insulating
regime of transport). In an intermediate field region ($\sim$10 - 30 V/cm
at 150 K or $\sim$5 - 15 V/cm at 77 K), a decreasing $I(E)$ characteristics
are realized, i.e. the NDC regime of response takes place. Both the
effective conductivity in the second Ohmic law region and the differential
conductivity under the NDC regime are increased with $l_c$.
\begin{figure}[ht]
\begin{center}
\includegraphics[scale=0.9]{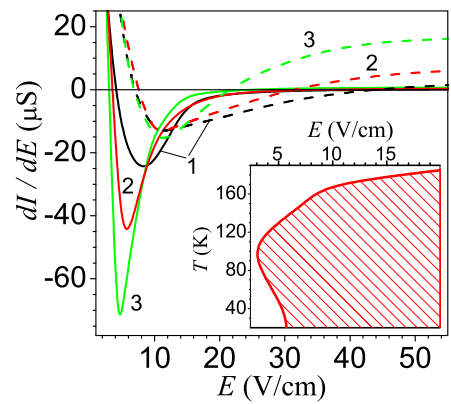}
\end{center}\addvspace{-1 cm}
\caption{(Color online) Differential conductivities $dI/dE$ at temperatures
$T=$150 and 77 K (dashed and solid curves) for $l_c=$5, 7.5, and 12.5 nm
(curves 1 - 3, respectively). Inset shows region of parameters, $E$ and $T$
(shaded) for which NDC regime takes place at $l_c=$7.5 nm. }
\end{figure}

In Fig. 10 we plot the differential conductivity $dI/dE$ versus $E$ for the
temperatures 150 and 77 K when the NDC regime takes place. Notice that
$dI/dE$ shows a strong dependency on $l_c$ in the low-temperature region
[c. f. Figs. 9(c) and 9(d)]. The NDC region over ($E$, $T$)-plane, where
$(dI/dE)<0$ is shaded in the inset for $l_c=$7.5 nm. A possibility for
the development of a spatial instability under NDC regime requires a special
investigation similar to the bulk case. \cite{17} But due to a small values
of $|dI/dE|$ ($\leq$100 $\mu$S at $T\geq$20 K) one may expect that an
inhomogeneous distribution is not developed for the typical size of samples.

\section{Conclusions}
Summarizing the consideration performed, we have examined the heating of
high-energy carriers taking into account the effective intra- and interband transitions caused by the optical phonons. The interplay of Joule heating
and recombination of electron-hole pairs gives rice to (i) the depletion of carriers' concentration and to (ii) the negative differential conductivity
at low temperatures. The model developed explains the nearly linear current-voltage characteristic at room temperature, in agreement with the experimental
data for intrinsic grapnene. \cite{7} Beside of this, temporal evolution of
response under an abrupt switching off a dc field lasts over a nanosecond
time scale, so that a direct measurement of transient response open up the way
to verify the relaxation and recombination mechanisms.

Further, we compare the consideration presented with the other results
published in order to stress that the nonlinear response is sensitive to a relative contribution of different relaxation processes and to a geometry
of measurements (size of sample, contacts). The last regime was considered
in short structures, \cite{3,4,5} see a recent data in Ref. 18 and a
description of the ballistic limit of transport was performed in Refs. 19.
In the case of monopolar transport \cite{6} the concentration is fixed and
$I(V)$ characteristics are only determined by the momentum and energy relaxation mechanisms, as it was discussed in Refs. 10 and 11.
Last but not least, the Joule heating is determined by the momentum scattering mechanism and it should be verified from the linear conductivity measurements,
see Fig. 1 and Refs. 7 and 12. An additional uncertainty appears due to the use
of any model of momentum relaxation because a microscopic mechanism remains
under debates. \cite{1,20}

Next, we discuss the assumptions used. The local time approach and the weak
anisotropy of distribution used in (2) are valid under the dominant momentum
scattering. We omitted the scattering by surface phonons of the substrate
in agreement with the consideration of the experimental data. \cite{7,21} If
such a contributions essential, it can shift the peculiarities considered (depletion of concentration and negative differential conductivity) to lower temperatures because the optical phonon energy $\sim$55 meV for the SiO$_2$ substrate. The carrier-carrier scattering is unessential here because the distributions obtained are spreaded over a range of energies up to
$\hbar\omega_{\Gamma , K}$, so that $f_p\ll 1$. Also one can neglect by the
long-range disorder \cite{22} because there is no low-energy particles if
$f_{p=0}\ll 1$. Since $v_d\gg v_{\Gamma , K}$, one can neglect the scattering
by optical phonons in comparison by the elastic scattering under consideration
of the momentum scattering. In addition, a heating of phonons is neglected
under the condition of effective phonon thermalization. \cite{23} Besides,
we are restricted by the quasiclassical approach, neglecting a mix of bands
due to a strong field (see \cite{18} and the quasiclassical condition in
Appendix of Ref. 9a), which can be essential in a lateral $pn$-junction.
The simplifications listed do not change either the character of the
high-field distributions or the numerical estimates for the current-voltage characteristics.

To conclude, both a detail theoretical consideration, including numerical
modeling for different scattering and recombination mechanisms, and an
experimental study of heating at low temperatures are timely now. An
observation of the peculiarities obtained and a verification of relaxation processes can be performed for a homogeneous samples with the four-point
geometry of contacts.

\end{document}